\documentclass{PoS}

\title{THE ARCADE RAMAN LIDAR SYSTEM FOR THE CHERENKOV TELESCOPE ARRAY}

\ShortTitle{ARCADE Lidar for the CTA}

\author{\speaker{L. Valore}$^{ab}$, C. Aramo$^b$, M. Doro$^c$, M. Iarlori$^d$, V. Rizi$^d$, A.S. Tonachini$^{ef}$,
P. Vallania$^{fg}$ for the CTA Consortium\footnote{Full consortium author list at http://cta-observatory.org}\\
\llap{$^a$} Università degli Studi di Napoli, Italy \\
\llap{$^b$} Sezione INFN di Napoli, Italy\\
\llap{$^c$} MPI for Physics, Munich and INFN Padova\\
\llap{$^d$} CETEMPS/DSFC Università degli Studi dell'Aquila, Italy\\
\llap{$^e$} Università di Torino, Italy\\
\llap{$^f$} Sezione INFN di Torino, Italy\\
\llap{$^g$} INAF Torino, Italy\\

E-mail: \email{valore@na.infn.it}}

\abstract{The Cherenkov Telescope Array (CTA) is the next generation of ground-based very high 
energy gamma-ray instruments; the facility will be organized in two arrays, one for each hemisphere. 
The atmospheric calibration of the CTA telescopes is a critical task. The atmosphere affects the 
measured Cherenkov yield in several ways: the air-shower development itself, the variation of the 
Cherenkov angle with altitude, the loss of photons due to scattering and absorption of Cherenkov 
light out of the camera field-of-view and the scattering of photons into the camera.  
In this scenario, aerosols are the most variable atmospheric component in time 
and space and therefore need a continuous monitoring. Lidars are among the most used instruments 
in atmospheric physics to measure the aerosol attenuation profiles of light. The ARCADE Lidar 
system is a very compact and portable Raman Lidar system that has been built within the FIRB 2010 
grant and is currently taking data in Lamar, Colorado. The  ARCADE Lidar is proposed to operate 
at the CTA sites with the goal of making a first survey of the aerosol conditions of the selected 
site and to use it as a calibrated benchmark for the other Lidars that will be installed on site.
 It is proposed for CTA that the ARCADE Lidar will be first upgraded in Italy and then tested in 
parallel to a Lidar of the EARLINET network in L'Aquila. Upgrades include the addition of the water 
vapour Raman channel to the receiver and the use of new and better performing electronics. It is 
proposed that the upgraded system will travel to and characterize both CTA sites, starting from the first 
selected site in 2016.}

\FullConference{The 34th International Cosmic Ray Conference,\\
		30 July- 6 August, 2015\\
		The Hague, The Netherlands}

\begin{document}

\section{Introduction}
The Cherenkov Telescope Array (CTA) will be the next generation of very high 
energy gamma-rays telescopes based on Imaging Astronomical Cherenkov 
Telescopes (IACTs). The observatory will consist of two arrays of telescopes,
one in the Northern and one in the Southern hemisphere to cover the full sky, 
each composed by several tens of IACTs of different size (Small Size Telescope - SST, Medium Size Telescope - MST,
Large Size Telescope - LST) to span a wide energy range going from a few tens of GeV up to more than 100 TeV.
The observatory will reach an unprecedented sensitivity, angular and energy resolution 
with respect to any of the existing gamma-rays observatories.

\begin{figure}[h!t]
   \centering
   \includegraphics[width=.8\textwidth]{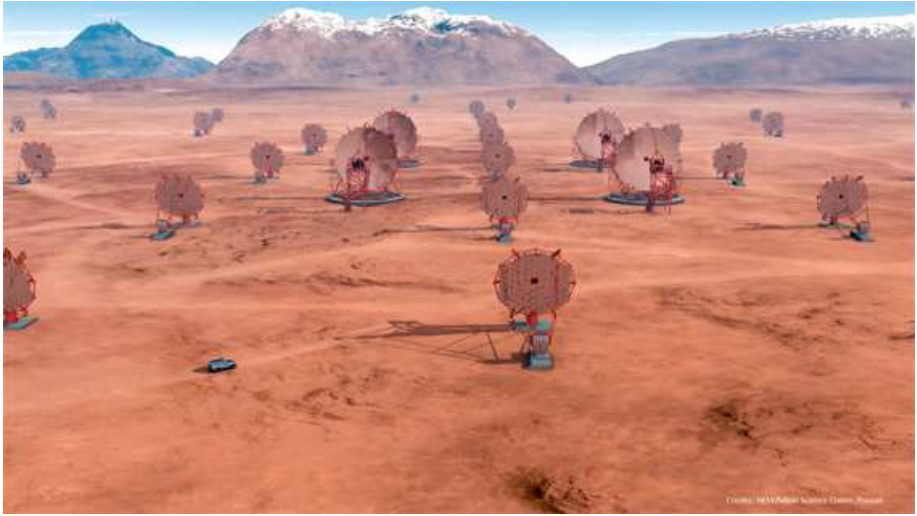}
  \caption{Sketch of a possible CTA site}
  \label{CTA}
 \end{figure}

%Primary gamma rays and cosmic rays entering the Earth's atmosphere generate
%an electromagnetic shower 
The telescopes will observe the Cherenkov light emitted by the electromagnetic
showers generated in the atmosphere due to the interaction of primary gamma rays 
with the atmosphere. The analysis of the showers images on different telescopes
allows the reconstruction of the primary particle energy and direction. 
The atmosphere therefore plays a double role, being responsible 
for both the production and the attenuation of the Cherenkov light from the 
point of emission to the telescopes.  Nowadays,
VHE gamma ray experiments include instruments to measure the atmospheric transparency
however mainly use their data to identify good quality observations time slots rather than to correct
data, even if some preliminary study is being done in this direction \cite{atmo_1,atmo_2,atmo_3,atmo_4,markus}. 
In the CTA, to ensure high quality data for the observatory, 
a continuous characterization of the local optical 
properties of the atmosphere during data taking will be performed to correct data \cite{markus_2}. 
Studies conducted using Monte Carlo simulations 
of the Magic telescopes proved the need for height-dependence aerosol attenuation 
measurements to correct IACT data \cite{garrido}: 
in case of aerosol over-density located close to the ground, the energy 
correction can be precisely obtained from the total transmission, otherwise altitude-resolving instruments
providing local aerosol attenuation profiles are needed.    
For this reason, the CTA comprises a multi-instrument atmospheric monitoring system \cite{Michele_atmohead}
that will include  Raman Lidars to measure the aerosol attenuation properties of the 
atmosphere, i.e. vertical aerosol optical depth profiles \cite{atmohead_1}. 
Briefly, the scattering of light by aerosols and molecules in the atmosphere can be divided into two general types: 
elastic scattering on both molecules and aerosols (in this case the wavelength of the scattered photon is the same 
of that of the laser source), and inelastic Raman scattering on a specific molecule (in this case the wavelength of the 
backscattered light is shifted). The intensity of the inelastic component of the backscattered light is significantly lower
($\sim$ 3 orders of magnitude) than the intensity of the elastically scattered light. The analysis of the backscattered
light allows to infer the aerosol attenuation conditions of the atmosphere above the observation point. 

The ARCADE Lidar, a Raman Lidar system built in Italy for the Atmospheric Research
for Climate and Astroparticle DEtection project \cite{icrc_2013, atmohead_2014, icrc_2015} 
and currently operating in Colorado, USA,
will be used to perform a first survey of the aerosol conditions on both CTA selected sites and as a 
benchmark to the other Raman Lidar systems that are scheduled to be running on site.  

\begin{figure}[h!t]
   \centering
   \includegraphics[width=.6\textwidth]{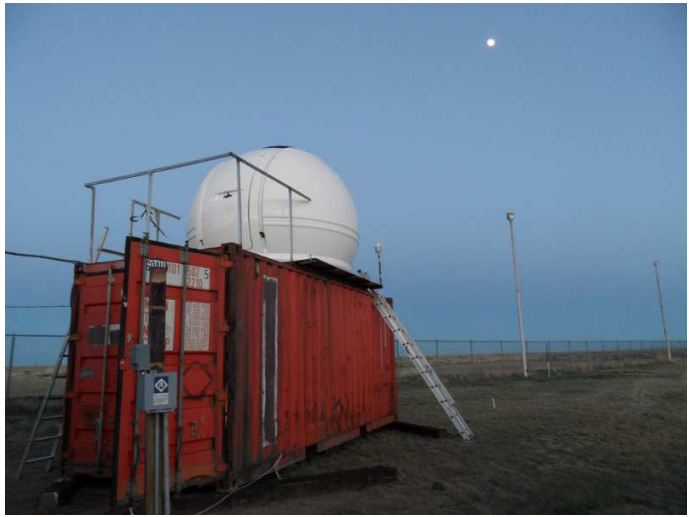}
  \caption{The ARCADE Lidar system in Lamar, Colorado}
  \label{foto_arcade}
 \end{figure}

\section{The ARCADE Raman Lidar}
The ARCADE Raman Lidar has been built at the University of Torino / INFN mechanical workshop of Torino 
in collaboration with the University of Naples / INFN Napoli within the FIRB 2010 grant funded by the
MIUR (Italian Ministry for Research), a 3 years project started in March 2012 aiming to characterize the
optical properties of the atmosphere, in particular measuring the aerosol attenuation profiles of UV light, for ground 
and space based cosmic rays experiments with the typical techniques used in this field :
side-scattering measurements using distant laser facilities and elastic / Raman lidars \cite{valore_icrc2009}. 
Data have been collected in Lamar, Colorado, since June 2014. The location was chosen for the arid-desertic environment,
similar to typical cosmic rays experiments sites.  It is a stand-alone atmospheric monitoring device
 not integrated in any ongoing experiment, therefore there has been no limitation on the data taking, allowing
 to compare for the first time the different techniques simultaneously on the same air mass with a good statistics. 
The ARCADE project has currently ended data taking and data analysis has just started. 
The details of the project can be found elsewhere in these proceedings \cite{icrc_2015}.  

The ARCADE Raman Lidar has been built to be very compact to ensure portability and a good stability of the 
mechanical structure to keep the alignment of the mirror with the laser beam and optics. For this reason,
the laser bench and the receiver are all assembled together and move at the same time on the zenith axis from 
0 to 90 degrees. This reduces possible misalignments and makes all the system easily portable. 
The computers and electronics are housed inside a 20 feet shipping container while the lidar is positioned on 
top of the container inside an astronomical dome that can be remotely operated, as the whole system. 
The system has been installed in Colorado during June 2014 and data have been collected for one year.

The ARCADE lidar is able to separate the elastic (355 nm) and nitrogen (387 nm) 
backscattered Raman components of the returning 
signal and to process them separately.  
Analysis methods that can be performed with the present configuration are the multi-angle analysis applied to the 
elastic traces and the Raman analysis on vertical and inclined shots, as described in \cite{lidar_paper}. 

\begin{figure}[h!t]
   \centering
   \includegraphics[width=.4\textwidth]{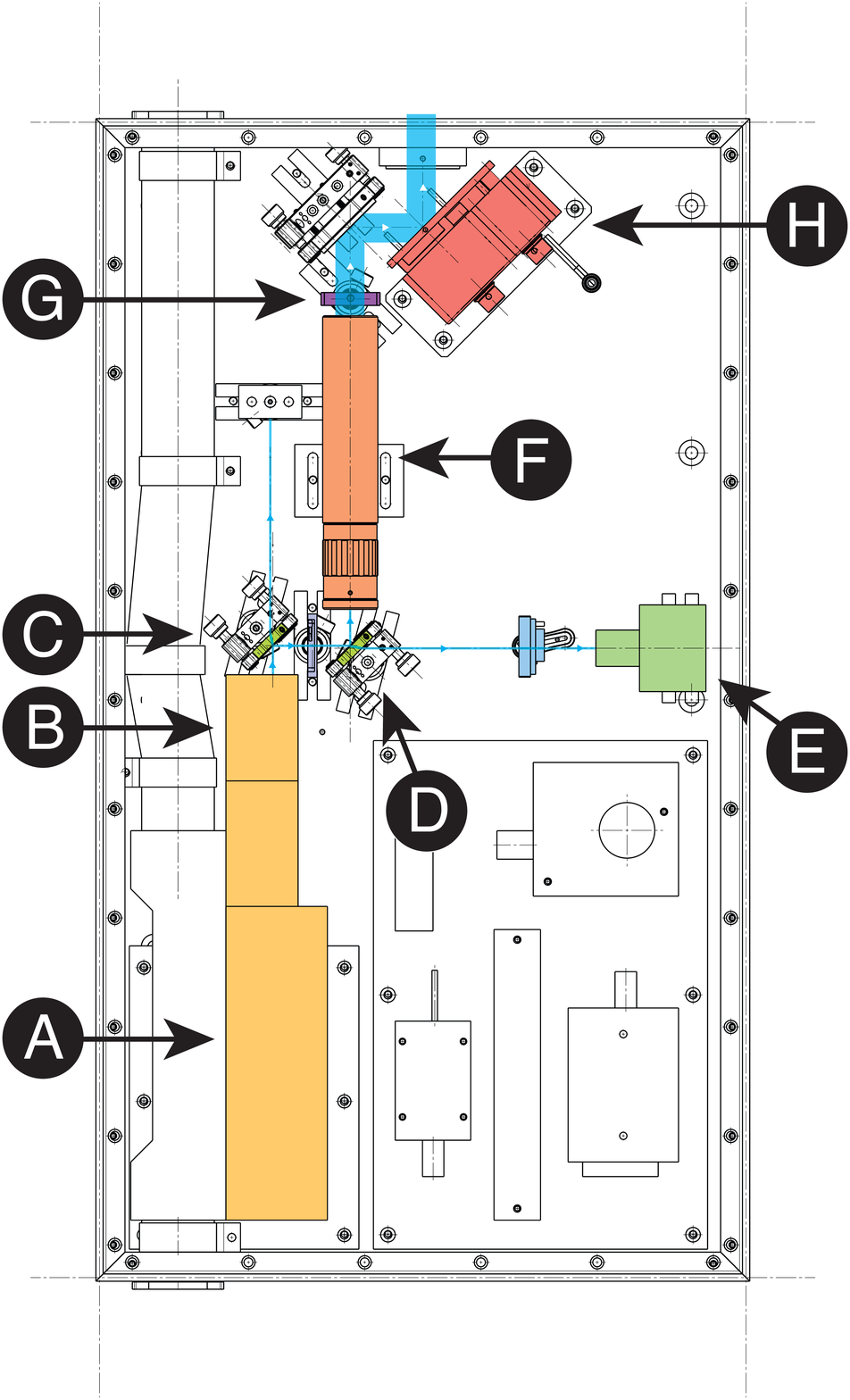}
   \caption{The design of the laser bench. A : the laser; B : high purity 4 dichroics module; C : fifth dichroic; D : 95/5 beam splitter; E : pyroelectric probe; F : 10X beam expander; G : depolarizer; H : motorized remotely controllable mirror.}
  \label{laser_bench}
 \end{figure}

The actual design of the ARCADE lidar includes a laser bench, an optical receiver and a mount for zenith steering 
in a very compact assembly. A sketch of the laser bench is shown in figure \ref{laser_bench}.

\begin{figure}
 \centering
 \includegraphics[width=0.5\textwidth]{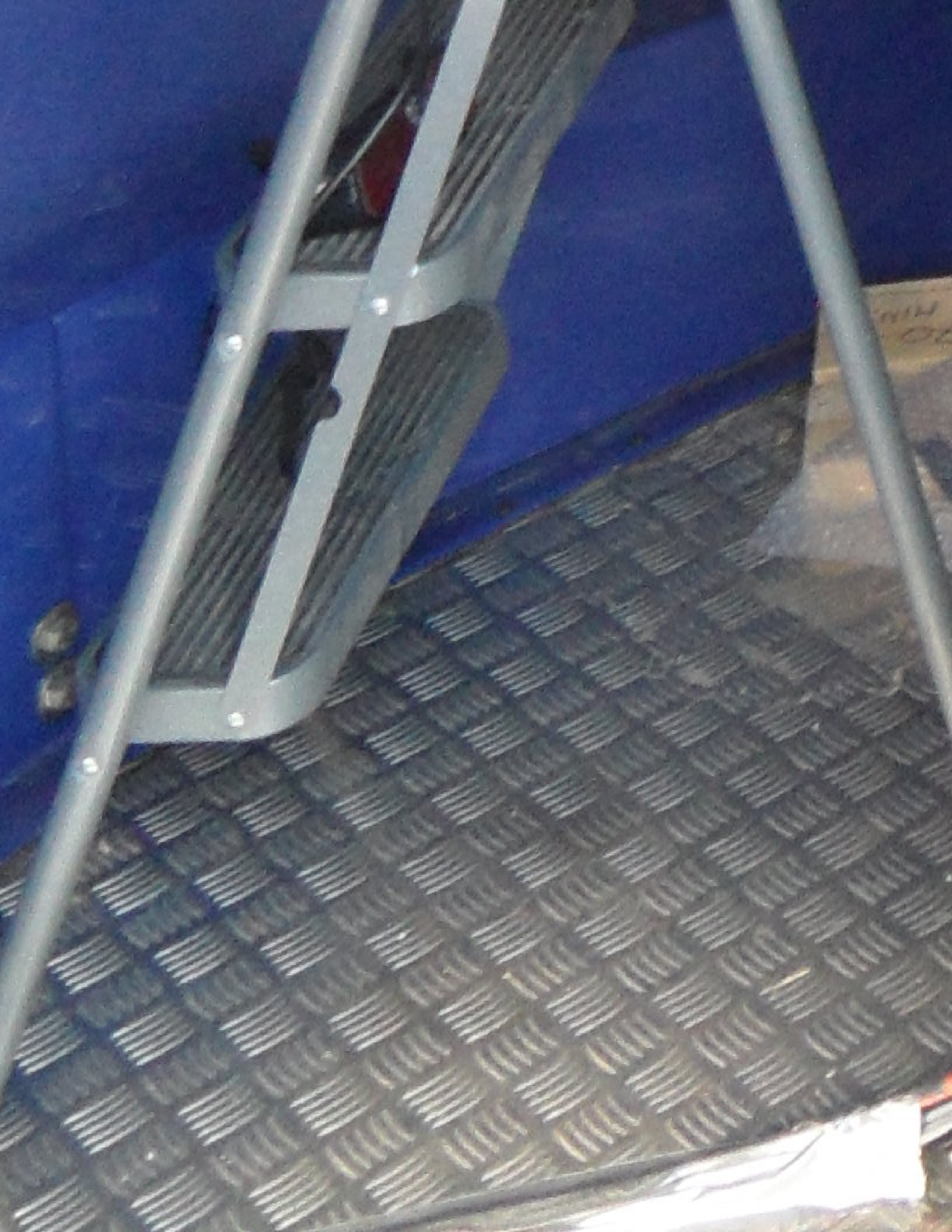}
 \caption{The ARCADE Lidar system. The black box includes the laser bench, primay mirror and receiver. The steering mechanism is visible. The system is located inside the astronomical dome.}
 \label{photo_lidar}
 \end{figure}

The laser source is a Quantel Centurion Nd:YAG firing at the maximum energy of 6 mJ at the third harmonic (355 nm). 
The first two harmonics (1064 nm and 532 nm) are suppressed by 5 consecutive dichroic mirrors to ensure a good purity 
of the laser beam. The laser energy is constantly monitored by means of a LaserProbe Inc. RjP-445 pyroelectric energy probe, 
which receives 5$\%$ of the beam output; the remaining 95$\%$ is sent to a 10X beam expander that reduced the laser beam 
divergence down to 0.3 mrad and enlarges the spot diameter to $\approx$ 20 mm. 
Before exiting the laser box, the beam is depolarized : this is due to the need of ensuring an isotropic
scattering of the UV laser light in air, for the side-scattering measurements with a distant telescope in the
ARCADE project.  A two-axis motorized mirror mount, holding a 2'' flat mirror, remotely controllable by computer, 
allows fine adjustments of the output beam direction : this is useful to verify (and adjust) the laser beam and 
mirror alignment for the lidar measurements.
The backscattered light is collected by a 25 cm diameter primary f/3 parabolic mirror. A flat secondary mirror 
sends the light onto a lens that defocuses the light beam. The light is thus guided onto a beam splitter which 
separates the elastic and N2 Raman backscatter photons. The resulting elastic beam passes through a 354.7 nm 
narrowband filter, while the Raman-shifted line from nitrogen (~R = 387 nm) is selected by a second narrowband filter. 
The two beams are read by two separate R1332Q photomultiplier tubes by Hamamatsu. 
Signals collected by the photomultipliers are then amplified and sampled with a 10 bit 2 GS/s CAEN digitizer \cite{caen}. 
The parallax between the laser beam and the receiver is 301 mm. 
A picture of the ARCADE lidar inside the astronomical dome is shown in figure \ref{photo_lidar}.

\begin{figure}[p!h!t]
 \centering
 \includegraphics[width=0.5\textwidth]{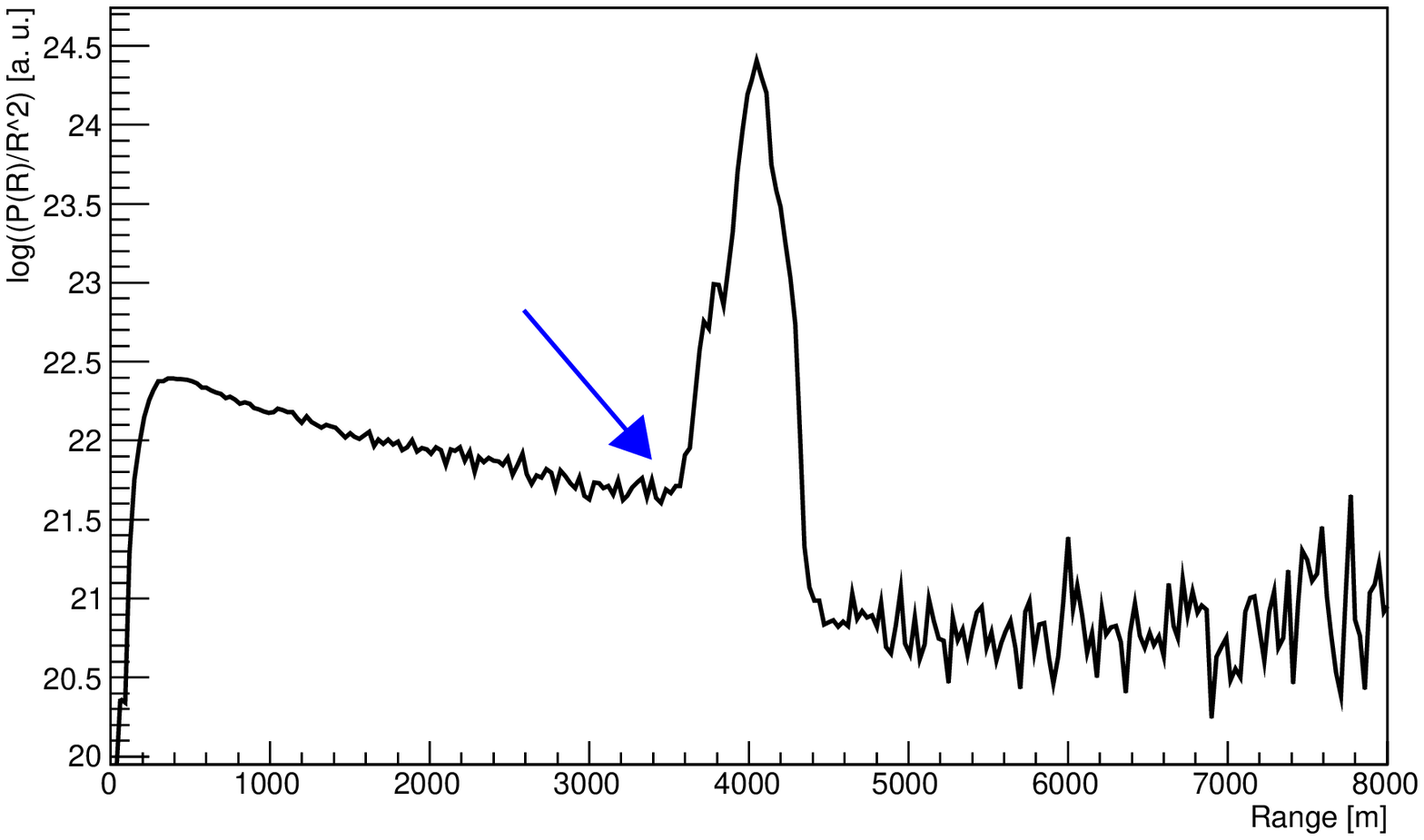}
 \includegraphics[width=0.5\textwidth]{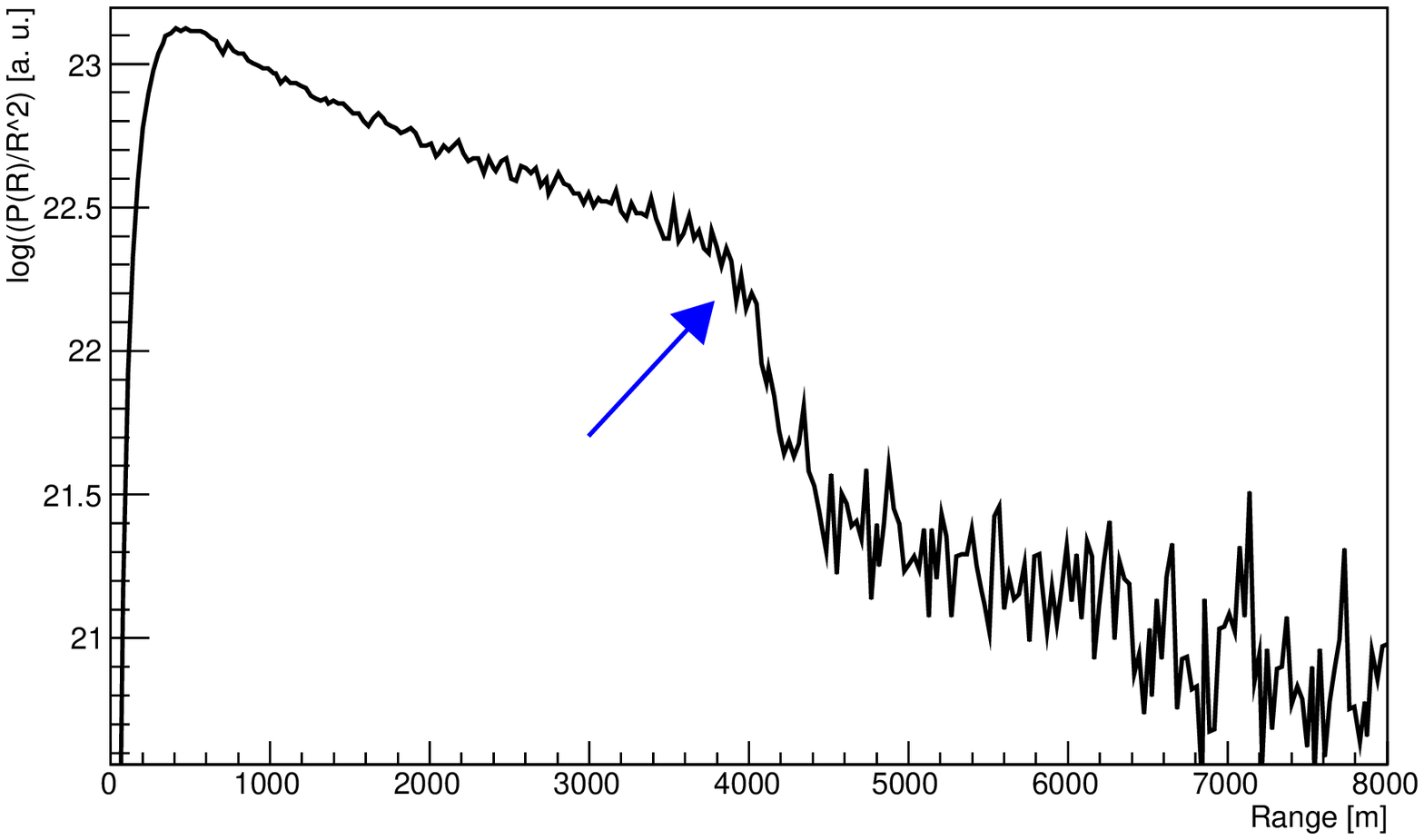}
 \caption{An example of a Raman and an elastic signals acquired by the ARCADE Lidar. A cloud is present above the lidar: it appears as a bump in the elastic signal (top) and as a drop in the Raman one (bottom). }
 \label{segnali_lidar}
 \end{figure}

With this configuration, the system can scan the atmosphere in a range from 250 m to 10 km. An example of one signal 
acquired by the Raman and elastic channel in presence of a cloud is shown in picture \ref{segnali_lidar}.

%\begin{figure}[h!t]
%   \centering
%   \includegraphics[width=.4\textwidth]{figures/.eps}
%  \caption{Elastic and nitrogen Raman signals as recorded by the ARCADE lidar. A cloud is present.}
%  \label{foto_laser_bench}
% \end{figure}

Data analysis has just started. The analysis of both the Raman signals (vertical and inclined) and of the 
elastic signals using the multiangle analysis will allow to measure the vertical aerosol optical depth profiles 
and to verify the horizontal uniformity of the atmosphere.

\section{The planned upgrade}

The ARCADE project has just ended and the lidar is planned to be shipped to the selected CTA sites to 
perform a survey of the atmosperic aerosol conditions on site and to be used as a benchmark for the first 
data acquisition of the other Lidar systems that will be operating on site. 
The lidar is unmounted in Lamar, Colorado in July 2015 to be shipped back to Torino, Italy for some optimization 
and upgrade before the transfer to the first CTA site.
 
The list of improvements will include : 

\begin{itemize}
\item new photomultipliers that will substitute the old ones, recovered from a previous experiment;
\item a new HV control system, more efficient and easily accessible (the present one is custom made);
\item a better insulation for the Lidar box (the Centurion laser suffers very low temperatures - a fan heater
controlled by a thermostat has been installed in the box during this winter, but it is not sufficient to ensure the
stability of the laser);
\item add the water vapour Raman channel to the receiver. This includes the modification of the receiver to
host a third photomultiplier and the related filters.
\end{itemize}

The addition of a second Raman channel at 407 nm will improve the precision of the measurements and will allow to 
obtain the water vapour profiles in atmosphere. 

The upgrade of the system will take place in Torino between September and December 2015. At the beginning of 2016
the upgraded Lidar will be shipped to L'Aquila to operate for some time in parallel with the Lidar of 
the DSFC / University of L'Aquila which is part of the EARLINET network and can be used as a benchmark to test the
performances of the new ARCADE lidar before its shipment to the selected site, that is predicted to take
place in the second half of 2016.

\section{The role of ARCADE in the CTA}

The ARCADE lidar will characterize both CTA sites. Presently it has not yet been defined which of the finally 
selected Northern and Southern site it will start with. A possibility may be the operation of 
the ARCADE lidar first at the Observatory del Roque de los Muchachos at La Palma, in the Canary Islands, where the Magic telescopes 
are currenlty running and where the first prototype of the Large Size Telescope for the CTA will be installed 
during 2016 together with other Raman lidars under development. 

The target of the ARCADE campaign in the CTA is to characterize the
seasonal (or climatological) aerosol content in the atmosphere over the site. ARCADE can measure the aerosol 
extinction $\rm \alpha(h)$ and volume backscatter $\rm \beta(h)$ coefficients profiles, as well as the water vapour
mixing ratio.  $\rm \alpha(h)$ and $\rm \beta(h)$ can fully describe the optical properties of the aerosols 
(i.e. the vertical aerosol optical depth and also a first guess of the aerosol scattering properties). 
The water vapour profiles, measured simultaneously to the aerosol attenuation profiles, 
can help to define the physical properties of the aerosol particles. ARCADE has also
steering capabilities from 0 to 90 degrees : the possibility to sample the atmosphere in inclined direction 
and/or horizontally is useful to investigate to which extent the atmosphere is horizontally homogeneous.

\end{document}